\documentclass[aps,twocolumn,prd,showpacs,preprintnumbers,nofootinbib]{revtex4}
\usepackage{amsmath}
\usepackage{graphicx}
\usepackage{subfigure}
\usepackage{dcolumn}
\usepackage{bm}
\usepackage{amssymb}
\usepackage{latexsym}
\usepackage{color}
\usepackage{graphicx}
\usepackage{dcolumn}
\usepackage{bm}

\newcommand{\be}{\begin{equation}}
\newcommand{\ee}{\end{equation}}
\newcommand{\bse}{\begin{subequations}}
\newcommand{\ese}{\end{subequations}}
\newcommand{\bea}{\begin{eqnarray}}
\newcommand{\eea}{\end{eqnarray}}

\begin{document}

\title{Phase transitions and perfectness of fluids in weakly coupled real
scalar field theories}
\author{Jiunn-Wei Chen$^{1,2}$, Mei Huang$^{3}$, Yen-Han Li$^{1}$, Eiji
Nakano$^{1}$, Di-Lun Yang$^{1}$}
\affiliation{$^{1}$ Department of Physics and Center for Theoretical Sciences, National
Taiwan University, Taipei 10617}
\affiliation{$^{2}$ CTP, Massachusetts Institute for Technology, Cambridge, MA 02139}
\affiliation{$^{3}$ Institute of High Energy Physics, Chinese Academy of Sciences,
Beijing 100049}

\begin{abstract}
We calculate the ratio $\eta /s$, the shear viscosity ($\eta )$ to entropy
density ($s)$, which characterizes how perfect a fluid is, in weakly coupled
real scalar field theories with different types of phase transitions. The
mean-field results of the $\eta /s$ behaviors agree with the empirical
observations in atomic and molecular systems such as H$_{2}$O, He, N, and
all the matters with data available in the NIST database. These behaviors
are expected to be the same in $N$ component scalar theories with an $O(N)$
symmetry. We speculate these $\eta /s$ behaviors are general properties of
fluid shared by QCD and cold atoms. Finally, we clarify some issues
regarding counterexamples of the conjectured universal bound $\eta /s\geq
1/4\pi $\ found in Refs. \cite{Cohen:2007qr,Dobado:2007tm}.
\end{abstract}

\maketitle




Quantum chromodynamics (QCD) is believed to undergo rapid transitions from
hadronic phases to a quark-gluon plasma (QGP) phase at high temperature $T$
and to quark matter phases at high quark chemical potential $\mu $ (see \cite%
{Rajagopal:2000wf,Stephanov:2007fk,Jacobs:2007dw}\ for reviews). Lattice
results show that the phase transition of the hadronic matter to QGP at
finite $T$ with $\mu =0$ is likely a crossover \cite{Lattice-QCD}. At finite 
$\mu $ and $T=0$, there is no reliable lattice result due to the severe
fermion sign problem. However, arguments based on a variety of models show
that the phase transition is of first-order. This first-order phase
transition turns into a crossover at smaller $\mu $ and finite $T$ at the
QCD critical end-point (CEP) \cite{Asakawa:1989bq}. There are lots of
interests in the CEP. Recently, it was proposed to probe the CEP by using
the ratio of shear viscosity $\eta $ to the entropy density $s$ of QCD \cite%
{Lacey:2006bc}.

Shear viscosity $\eta $ characterizes how strongly particles interact and
move collectively in a many-body system. In general, the stronger the
interparticle interaction, the smaller the shear viscosity (here $\eta $ is
normalized by the density). It was conjectured \cite{KOVT1} that no matter
how strong the particle interaction is, $\eta /s$ has a universal minimum
bound $1/4\pi $ in any system. This bound is motivated by the uncertainty
principle and is found to be saturated for a large class of strongly
interacting quantum field theories whose dual descriptions in string theory
involve black holes in anti-de Sitter space \cite{KOVT1}.

There are two important questions regarding this conjecture. The first one
is: Is this $\eta /s$ bound truly universal? By definition, there is no
proof of this conjecture yet. From experimental observations, the bound is
well satisfied in matters like H$_{2}$O, N and superfluid He \cite{KOVT1}.
For cold fermionic atoms with an infinite scattering length (the unitarity
limit), the bound is satisfied but $\eta /s$ is close to the bound near the
phase transition temperature $T_{c}$ \cite{etas-supfluid}. Similar behavior
is found for QCD at $\mu =0$ \cite{Csernai:2006zz,Chen:2006iga}.
Relativistic heavy ion collisions (RHIC) \cite%
{RHIC,Molnar:2001ux,Teaney:2003pb,Romatschke:2007mq} and lattice
computations of a gluon plasma \cite{etas-gluon-lat}) suggest that $\eta /s$
of QCD is close to the minimum bound at just above $T_{c}$. But one cannot
conclude whether the bound is violated based on current precision. On the
other hand, interesting counterexamples of the bound have been constructed 
\cite{Cohen:2007qr,Dobado:2007tm} and a possible modification of the bound
due to some string excitations in the dual theory \cite%
{Kats:2007mq,Brigante:2007nu,Ge:2008ni}.

The second question is: What is the general behavior of $\eta /s$,
especially for QCD and cold atom systems which are of high interests but not
well known? The answer to this question is interesting on its own. It also
has interesting applications. For example, locating QCD\ CEP by $\eta /s$
requires this information. Also, knowing when a system will reach its
minimum $\eta /s$ is important for testing the minimum bound conjecture. By
now it is known that several different systems have qualitatively the same $%
\eta /s$ behaviors. For example, H$_{2}$O, N, and He all have the minimum $%
\eta /s$ near the liquid-gas phase transition temperature. The same is true
with QCD at $\mu =0$ \cite{Csernai:2006zz,Chen:2006iga} and near the nuclear
liquid-gas phase transition \cite{Chen:2007xe}. It is also true in cold
atoms in the unitarity limit \cite{etas-supfluid}. It would certainly be
interesting to further map out the detailed $\eta /s$ structure in the phase
diagram of QCD and other systems.

In this letter, we address the above two questions at the same time. We
study how $\eta /s$ behaves in the simplest field theory---a real scalar
field theory. The resulting $\eta /s$ in first-, second-order phase
transitions and crossover behaves the same way as in H$_{2}$O, N, He, and
all the matters with data available in the NIST database \cite%
{webbook,Csernai:2006zz,Chen:2007xe}. We argue that this agreement might
hold when the theory is generalized to $N$ components with an $O(N)$
symmetry, which is the low energy effective field theory of a big class of
systems. We then speculate these behaviors might be general properties of
fluid which are shared by QCD and cold atoms. Furthermore, this simple
theory allows us to clarify some issues regarding the counterexamples of the 
$\eta /s$\ bound \cite{Cohen:2007qr,Dobado:2007tm} in a more transparent way.

We will study a real scalar theory with the Lagrangian%
\begin{equation}
\mathcal{L}=\frac{1}{2}(\partial _{\mu }\phi )^{2}-\frac{1}{2}a\phi ^{2}-%
\frac{1}{4}b\phi ^{4}-\frac{1}{6}c\phi ^{6}.
\end{equation}%
This theory is invariant under $\phi \rightarrow -\phi $ and has a $Z_{2}$
symmetry. There could be two additional terms with dimension six: $\phi
^{3}\partial ^{2}\phi $, $\phi \partial ^{2}\partial ^{2}\phi $ (the other
terms are related to these ones by integration by parts). These terms can be
removed by field redefinition or, equivalently, by applying the equation of
motion. The inclusion of the dimension six terms shows that this is an
effective field theory, which is valid under the cut-off scale $1/\sqrt{c}$
and is renormalized order by order in the momentum expansion $p\sqrt{c}$, $p$
being a typical momentum scale in the problem. $a$, $b$, and $c$ are
renormalized quantities and the counterterm Lagrangian is not shown. The
renormalization condition is that the counterterms do not change the
particle mass and the four- and six-point couplings at threshold. We will
discuss the following cases: 1) $c=0,$ $b>0,$ $a>0$, the system is always in
the symmetric phase. 2) $c=0,$ $b>0,$ $a<0$, the vacuum at $T=0$ breaks the $%
Z_{2}$ symmetry spontaneously. However, the symmetry is restored at higher $%
T $ with a second-order phase transition. 3) Adding an explicit symmetry
breaking term $\delta L=H\phi $ to the Lagrangian of 2) to model a
crossover. 4) $c>0,$ $b<0,$ $a>0$, the broken symmetry is restored at high $%
T $ with a first-order phase transition.

We will focus on the case of weak coupling and compute the mean-field
effective potential via the standard Cornwall--Jackiw--Tomboulis (CJT)
formalism \cite{CJT} which has the one-particle irreducible diagrams
included self-consistently. The effective potential in the CJT formalism
reads \cite{Lenaghan:1999si} 
\begin{eqnarray}
V[\bar{\phi},S] &=&\frac{1}{2}\int_{K}\left[ \,\ln
S^{-1}(K)+S_{0}^{-1}(K)\,S(K)-1\,\right] \newline
\notag \\
&&+\,\,V_{2}[\bar{\phi},S]+U(\bar{\phi})\,\,,
\end{eqnarray}%
where $U(\bar{\phi})=a/2~\bar{\phi}^{2}+b/4~\bar{\phi}^{4}+c/6~\bar{\phi}%
^{6} $ is the tree-level potential, and $S(S_{0})$ is the full(tree-level)
propagator: 
\begin{equation}
S^{-1}(K,\bar{\phi})=-K^{2}+m^{2}(\bar{\phi})\;,\newline
S_{0}^{-1}(K,\bar{\phi})=-K^{2}+m_{0}^{2}(\bar{\phi})\;,
\end{equation}%
with the tree-level mass $m_{0}^{2}=a+3b~\bar{\phi}^{2}+5c~\bar{\phi}^{4}$.

In this work, we neglect the non-tadpole type loop diagrams. This
\textquotedblleft Hartree approximation\textquotedblright\ is good when $%
T\gg \left\vert b\right\vert ^{1/2}T_{c}$ for $c=0$ and when $T\gg
\left\vert a\right\vert ^{1/2}$ for $c>0$. Thus, to the order we are
working, the 2 PI potential $V_{2}$ only includes 
\begin{equation}
V_{2}[\bar{\phi},S]=\left( \frac{3b}{4}+\frac{15c}{2}\bar{\phi}^{2}\right)
L\left( \bar{\phi}\right) ^{2}+\frac{15}{6}cL\left( \bar{\phi}\right) ^{3},
\label{V2}
\end{equation}%
where $L\left( \bar{\phi}\right) =\int_{K}\,S(K,\bar{\phi})$. The
self-consistent one- and two-point Green's functions satisfy 
\begin{equation}
\left. \frac{\delta V}{\delta \bar{\phi}}\right\vert _{\bar{\phi}=\phi
_{0},S=S(\phi _{0})}\equiv 0\;,\;\;\;\left. \frac{\delta V}{\delta S}%
\right\vert _{\bar{\phi}=\phi _{0},S=S(\phi _{0})}\equiv 0\;\;\;\;.
\end{equation}%
This allows us to solve $\phi _{0}$ and $m$ through the coupled equations: 
\begin{gather}
\phi _{0}\left( a+b\phi _{0}^{2}+c\phi _{0}^{4}+(3b+10c\phi _{0}^{2})L(\phi
_{0})+15cL(\phi _{0})^{2}\right) =0\,,  \notag \\
m^{2}-m_{0}^{2}=3(b+10c\phi _{0}^{2})L(\phi _{0})+15cL(\phi _{0})^{2}.\,
\end{gather}

The entropy density of the system is given as $s=-\partial V(\phi
_{0})/\partial T,$ while the shear viscosity $\eta $ is calculated using the
Boltzmann equation. It is proven that in a weakly coupled scalar field
theory with quartic and cubic terms, summing the leading order diagrams for $%
\eta $\ is equivalent to solving the Boltzmann equation with effective $T$
dependent mass and scattering amplitudes \cite{Jeon}. Thus, one can directly
apply the Boltzmann equation to compute $\eta $ for cases with $c=0$ for
both the symmetric ($a>0$) and symmetry breaking ($a<0$) cases. Furthermore,
since the proof of Ref. \cite{Jeon}\ does not use properties that are
restricted to scalar theories, the conclusion is expected to hold for more
general theories with weak couplings, including QCD in the perturbative
regime \cite{Arnold:2003zc}. Here, we also apply it to the $c>0$ case with a
first-order phase transition.

The two-particle elastic scattering amplitude, which governs particle
collisions in the Boltzmann equation, is%
\begin{equation}
i\mathcal{T}=\lambda _{4}+\lambda _{3}^{2}\left[ \frac{1}{s-m^{2}}+\frac{1}{%
t-m^{2}}+\frac{1}{u-m^{2}}\right] ,  \label{sca}
\end{equation}%
where $s,t$ and $u$ are Mandelstam variables, and $\lambda _{3}=6\phi _{0}(b+%
\frac{10c}{3}\phi _{0}^{2}+10cL\left( \phi _{0}\right) )$ and $\lambda
_{4}=12(\frac{b}{2}+5c\phi _{0}^{2}+5cL\left( \phi _{0}\right) )$ are
effective couplings.

\begin{figure}[tbp]
\par
\begin{center}
\includegraphics[width=7.5cm]{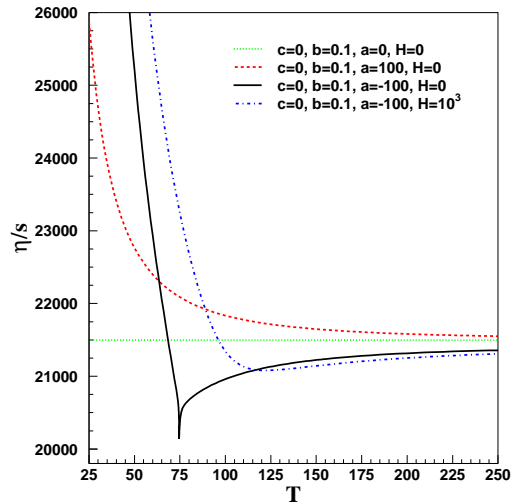} 
\end{center}
\par
\caption{$\protect\eta /s$ vs. $T$ for cases with a second-order phase
transition\ (solid curve), a crossover (dash-dotted curve), and with\ no
phase transition for massive field (dashed curve) and massless field (dotted
curve). Parameters can be in arbitrary units.}
\label{fig-2nd}
\end{figure}

\begin{figure}[tbp]
\par
\begin{center}
\includegraphics[width=7.5cm]{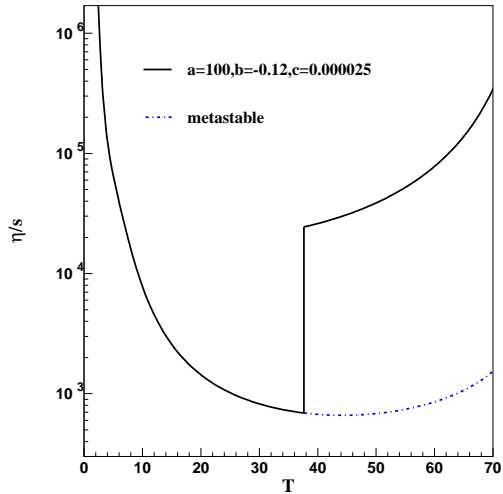} 
\end{center}
\par
\caption{$\protect\eta /s$ vs. $T$ for the ground state\ (solid curve) and
metastable state (dash-dotted curve) in a first-order phase transition.
Parameters can be in arbitrary units.}
\label{fig-1st}
\end{figure}

Now we discuss the behavior of $\eta /s$, starting from the simplest case: $%
a=c=0$ and $b>0$. In this theory, the only scale in the problem is $T$ and
the system is always in the symmetric phase. Thus, both $\eta $ and $s$ are
proportional to $T^{3}$ on dimensional ground and $\eta /s$ is $T$
independent (up to the logarithmic running of $b$ which we neglected).

When the mass term is added, $\eta /s$ is no longer a constant. $\eta /s$
curves for $b>0$, $c=0$ and (i) $a>0$ (ii) $a<0$ are shown in Fig. 1. (We
have used $b=0.1$. The expansion parameter is $\sim \lambda /(4\pi )^{2}$,
where $\lambda =6b$. \cite{Cohen:1997rt}) For (i), there is no phase
transition. $\eta /s$ is always monotonically decreasing. For (ii), $\eta /s$
reaches its minimum and develops a cusp at the $T_{c}$ of the second-order
phase transition. There is clearly a qualitative difference in the $\eta /s$
behavior between cases with and without a phase transition. In fact, in the
high $T$ expansion ($bT^{2}\gg \left\vert a\right\vert $) , 
\begin{equation}
\frac{\eta }{s}=\frac{k}{b^{2}}\left( 1+\frac{0.24}{\sqrt{b}}\frac{a}{T^{2}}%
+\cdots \right) \text{,}  \label{highT}
\end{equation}%
with $k=192+49\sqrt{b}+\mathcal{O}(b)$, consistent with $\eta $ computed in 
\cite{Jeon,Moore:2007ib}. Thus, the sign of $a$ determines whether $\eta /s$
is decreasing or increasing at high $T$. As $T\rightarrow 0$, $s$ approaches
zero exponentially (the excitations are massive) while $\eta $ approaches
zero by power laws. Thus, $\eta /s$ is decreasing in both cases at low $T$.
(This feature is not affected by the Hartree approximation used at low $T$.)
One concludes that $\eta /s$ is not monotonic in $T$ for a second-order
phase transition. Although our mean-field result would be modified very
close to $T_{c}$ (with $\left\vert T-T_{c}\right\vert /T_{c}\lesssim O(b)$),
we argue that the most natural scenario for $\eta /s$ in a second-order
phase transition is to have a single local minimum at $T_{c}$ (where $m$
also reaches its minimum, $m=0$), because there is no other $T$ that is more
special than the others to develop another minimum in $\eta /s$.
Analogously, the most natural scenario for $\eta /s$ with no phase
transition is monotonic decreasing because no $T$ is more special than the
others to develop a minimum. Finally, for a second-order phase transition,
one naturally expects a cusp at $T_{c}$, while the cusp is smoothed out in a
crossover ($H\neq 0$). These \textquotedblleft
naturalness\textquotedblright\ arguments do not depend on whether the
coupling is weak. In fact, there is no $b$ that is more special than the
others such that one does not expect the $\eta /s$ behavior to change in the
strong coupling cases ($\left\vert b\right\vert \gg 1$) either.

For the first-order phase transition ($a>0$, $b<0$, and $c>0$), the result
for $\eta /s$ is shown in Fig. 2. There is a discontinuity at $T_{c}$, as
expected. The $\eta /s$ is concave in general and its minimum is reached at $%
T_{c}$ from the low $T$ side. The qualitative behaviors above and below $%
T_{c}$ are similar to those of a second-order phase transition. The main
difference is the discontinuity. It is not intuitively obvious why this
should be the case because even though the effective potentials of the
first- and second-order phase transitions look similar near the minima (thus
the masses behave similarly) the Lagrangians are still quite different. It
turns out this is simply because the scattering amplitude of Eq.(\ref{sca})
is dominated by the $\lambda _{4}$ term in both cases as long as $%
T^{2}>\left\vert a\right\vert $. Also, the fact that the minimum of $\eta /s$
is reached at $T_{c}$ from below, instead of above, is physical but
nontrivial. It is physical, because particle interaction is stronger in the
low temperature side. It is nontrivial, because the discontinuities in $\eta 
$ and $s$ are of the same sign. So it is not obvious the discontinuity of $%
\eta /s$ has to be governed by $\eta $.

The above-mentioned behaviors of $\eta /s$ in first-, second-order phase
transitions and crossover are the same as those in H$_{2}$O, N, He, and all
the matters with data available in the NIST database \cite%
{webbook,Csernai:2006zz,Chen:2007xe}. We expect the agreement still holds
when the scalar field is generalized to $N$ components and the Lagrangian
has an $O(N)$ symmetry. This is because the arguments used to explain the $%
\eta /s$ behaviors do not depend on $N$, except for Eq.(\ref{highT}) in the $%
c=0$ case. However, in the $O(N)$ theory, the prefactors in Eq.(\ref{highT})
change \cite{Aarts:2004sd}\ but the signs remain. Thus, the arguments still
hold.

Given the general agreement in the $\eta /s$\ behaviors of H$_{2}$O, He, N, $%
O(N)$\ scalar field theories, and QCD and cold atoms in some known cases, it
is conceivable that the $\eta /s$\ behaviors shown in Figs. 1 and 2 are
general properties of a large class of fluids, including QCD and cold atoms.
If this is correct, then the best place to measure the minimum $\eta /s$ to
test the minimum bound conjecture is near $T_{c}$. In particular, for a pure
gluon theory which has a first-order phase transition, the minimum $\eta /s$
could be just below $T_{c}$. This strongly motivates future lattice
computations to be carried out in the lower $T$ regime. In addition, this
also provides theoretical support for locating QCD\ CEP by measuring $\eta
/s $. There is some subtlety regarding the CEP. In H$_{2}$O, He, or N, a
cusp in $\eta /s$ is observed at the CEP with experimental resolution.
Theoretically, it is argued that $\eta $ has a mild divergence (and hence $%
\eta /s$) at QCD\ CEP \cite{Son:2004iv} as in the Model $H$ according to
Hohenberg and Halperin's classification \cite{Hohenberg:1977ym} (but see 
\cite{Ohnishi:2004eb} for a caveat). Experimentally, divergence of $\eta $
in $^{3}$He liquid-gas transition critical point was observed \cite{He1} but
not for the tricritical point of $^{3}$He-$^{4}$He binary fluid \cite{He2}.
Both cases belong to Model $H$ and were predicted to have the same weak
divergence by Ref. \cite{Hohenberg:1977ym}. In any case, it is unlikely that
RHIC has the sensitivity to this effect.

Finally, we comment on the counterexamples of the minimum $\eta /s$\ bound
conjecture constructed in \cite{Cohen:2007qr,Dobado:2007tm}. A common
feature of those counterexamples is a large number of flavor $g$. In the
weak coupling and small density limits, one can compute $\eta $\ (also using
the Boltzmann equation) and $s$\ reliably. In the non-relativistic examples
employed in \cite{Cohen:2007qr,Dobado:2007tm}, particle numbers are fixed or
controlled by chemical potentials. It was found that $\eta $\ does not scale
with $g$\ (because it is normalized to the density already) but $s$\ does.
Thus, $\eta /s\propto 1/\left( Log(g)a_{s}^{2}\right) $\ as $g\rightarrow
\infty $ ($a_{s}$ is the two-body scattering length) and the $1/4\pi $\
bound was violated. There is an important difference in those
counterexamples on whether flavor changing is allowed during the collisions.
In the $O(N)$\ scalar field theory, $\phi _{1}\phi _{1}\rightarrow \phi
_{i}\phi _{i}$\ ($i=1,2...N$) collision can happen. Thus, it is a model with
flavor changing. To the leading order in the $b$\ expansion, $\eta /s\propto
1/(Nb^{2})$\ for $a=c=0$, similar to Eq.(\ref{highT}) but with an extra
factor of $N$\ from $s$. Naively the conjecture is violated when $%
N\rightarrow \infty $. However, the expansion parameter is $\left( Nb\right) 
$. One needs $b\propto 1/N$\ or smaller (and hence $\eta /s\propto N$\ or
larger) to be consistent with the weak coupling treatment. Thus, one cannot
conclude that the $O(N)$\ scalar field theory violates the bound in the
large $N$\ limit based on weak coupling argument. The same is true to the
non-relativistic flavoring changing counterexamples of \cite%
{Cohen:2007qr,Dobado:2007tm} ($a_{s}\propto 1/g$ is required). In the strong
coupling regime, the $b\propto 1/N$\ scaling seems not necessary. However,
without the scaling, the $\beta $\ function of $b$\ blows up in the large $N$%
\ limit. It is not clear whether a sensible non-perturbative theory can be
constructed this way. Our example sets no constraint on flavor conserving
cases in \cite{Cohen:2007qr,Dobado:2007tm}, though.

In Ref. \cite{Cohen:2007qr}, there is discussion about whether the $\eta /s$%
\ bound conjecture should be applied to metastable states. In the definition
using Kubo's formula, $\eta $\ is related to the linear response of an
ensemble. It is hard to exclude the ground state from the ensemble to
compute the metastable state $\eta $. A similar problem happens in computing 
$\eta $\ with the Boltzmann equation in which the dissipation of
perturbations away from thermal equilibrium gives rise to $\eta $. The
ground state information is encoded in the equilibrium distribution.
However, experimentally it makes perfect sense to measure $\eta $\ of a
material as long as it is stable during the time of the measurement. Water
is a sharp example of this \cite{Cohen:2007qr}. In electroweak interaction,
water is metastable to proton fusion of the two hydrogen atoms despite the
large Coulomb barrier. However, one can modify the Lagrangian by sending the
weak boson masses to infinity to stabilize water. The resulting $\eta $\
will be close to the experimentally measured value. It is not clear whether
the same result can be obtained by simply discarding near ground states from
the ensemble. Assuming this is correct, we investigate its consequence in
our model by expanding $\phi $\ around the false vacuum above the $T_{c}$\
of a first-order phase transition. As shown in Fig. 2, $\eta /s$\ of the
metastable state can be lower than the minimum $\eta /s$\ of the true ground
state (the effect will be more pronounced if $c$ is bigger). The metastable
state $\eta /s$\ has the tendency to keep decreasing until this state
eventually disappears at higher $T$. It would be interesting to see whether
this behavior persists in the strongly interacting theory.


We thank Tom Cohen, Yusuke Nishida, Dirk Rischke, Dam Son and Larry Yaffe
for useful discussions. MH thanks the National Taiwan U. for hospitality.
JWC thanks the U. of Washington for hospitality. This work is supported by
the IHEP, Chinese Academy of Sciences, CAS key project KJCX3-SYW-N2,
NSFC10735040, and the NSC and NCTS of Taiwan.


\end{document}